\documentclass[conference]{IEEEtran}
\IEEEoverridecommandlockouts
\usepackage{cite}
\usepackage{amsmath,amssymb,amsfonts}
\usepackage{graphicx}
\usepackage{textcomp}
\usepackage{xcolor}
\usepackage{lineno}
\usepackage{scalerel}
\usepackage{amssymb}
\usepackage{subfigure}
\usepackage{algorithm}
\usepackage{algorithmicx}
\usepackage{algpseudocode}
\usepackage{graphicx}

\def\BibTeX{{\rm B\kern-.05em{\sc i\kern-.025em b}\kern-.08em
    T\kern-.1667em\lower.7ex\hbox{E}\kern-.125emX}}
\begin{document}

\title{Collaborative Edge Caching: a Meta Reinforcement Learning Approach with Edge Sampling
\thanks{Corresponding author: Zhi Wang, wangzhi@sz.tsinghua.edu.cn. The first two authors contributed equally to this work. 
This research was funded by Shenzhen Science and Technology Program (Grant No. RCYX20200714114523079 and JCYJ20220818101014030). We would like to thank Kuaishou for sponsoring the research.}
}

\author{\IEEEauthorblockN{Bowei He$^\S$, Yinan Mao$^*$, Shiji Zhou$^\dagger$, Chen Ma$^\S$, Zhi Wang$^{*\dagger}$}
\IEEEauthorblockA{\textit{$\S$ Department of Computer Science, City University of Hong Kong, Hong Kong SAR} \\
\textit{$*$ Tsinghua Shenzhen International Graduate School, Tsinghua University, Shenzhen, China} \\
\textit{$\dagger$ Tsinghua-Berkeley Shenzhen Institute, Tsinghua University, Shenzhen, China} }
}

\maketitle

\begin{abstract}
Current learning-based edge caching schemes usually suffer from dynamic content popularity, e.g., in the emerging short video platforms, users' request patterns shift significantly over time and across different edges.
An intuitive solution for a specific local edge cache is to collect more request histories from other edge caches. 
However, uniformly merging these request histories may not perform satisfactorily due to heterogeneous content distributions on different edges. 
To solve this problem, we propose a collaborative edge caching framework. First, we design a meta-learning-based collaborative strategy to guarantee that the local model can timely meet the continually changing content popularity. Then, we design an edge sampling method to select more ``valuable'' neighbor edges to participate in the local training. To evaluate the proposed framework, we conduct trace-driven experiments to demonstrate the effectiveness of our design: it improves the average cache hit rate by up to $10.12\%$ (normalized) compared with other baselines.
\end{abstract}


\section{Introduction}
\label{sec:intro}
The emergence of content-rich applications has seen a surge in the amount of short-video content being delivered over the Internet globally. 
According to \cite{10.1145/3339825.3391856}, in 2022, the whole traffic for short videos had tripled the amount presented in 2021. To keep up with this demand, the caching system needs to be shifted from the cloud servers or the content delivery network to edge devices, bringing the contents closer to end viewers and helping reduce network congestion~\cite{9596584}. 
For instance, platforms such as TikTok and YouTube, which have experienced flash-increasing viewers due to user-generated contents, now can store the contents more efficiently in nearby edge caching devices than remote servers. 
However, the content request distribution (known as the content popularity) of edge contents is more volatile than that of cloud contents \cite{7128393,9148214}.

The evolution of the content request distribution dynamics has demonstrated the limitations of traditional rule-based caching methods such as Least Recently Used (LRU) and Least Frequently Used (LFU), among others~\cite{10.1145/2517349.2522722,zhou2021caching}. 
These methods rely on a basic stationary assumption concerning request patterns, e.g., considering past request frequencies or the time since the last request. 
To ensure their effectiveness, they must be applied to large groups of users; however, edge caches usually serve only a small user base, making these traditional methods inapplicable to edge content delivery.

\begin{figure*}[!t]
	\centering
\begin{minipage}[b]{0.99\linewidth}
    \centerline{\includegraphics[width=0.99\linewidth]{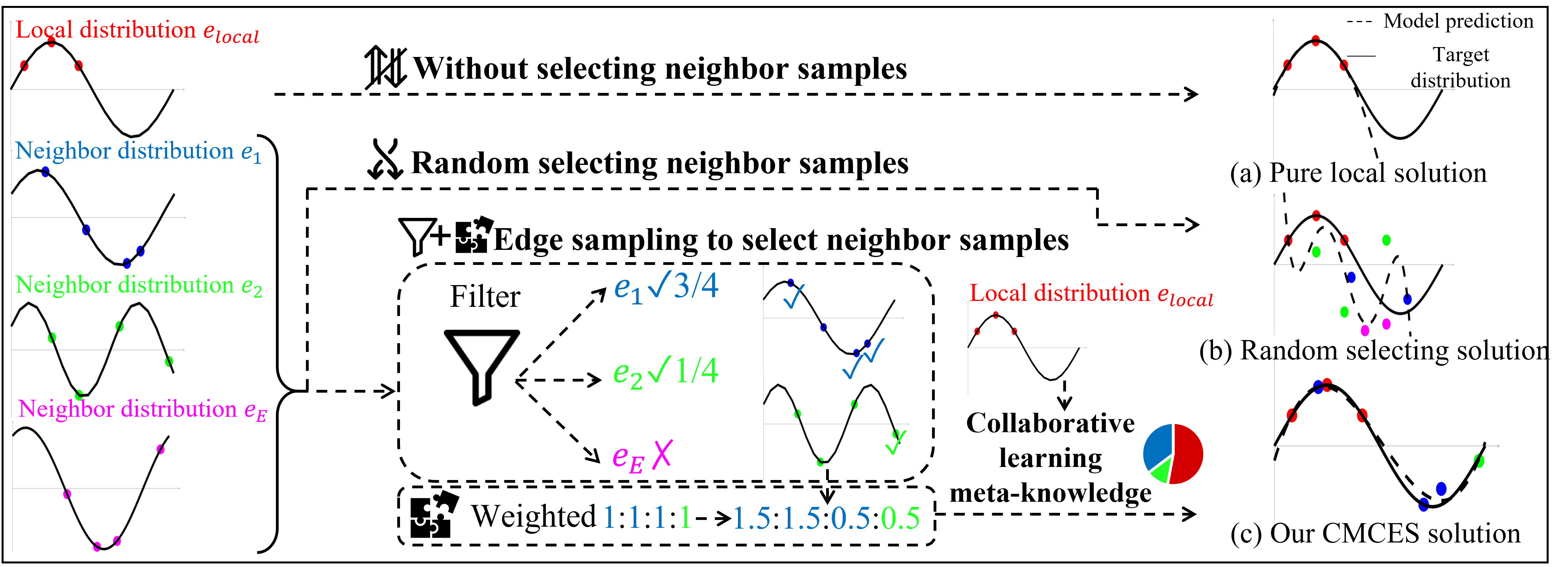}}
\end{minipage}
	\caption{\textbf{(a) Pure local solution} trained on limited samples fails to fit the target distribution $e_{local}$; \textbf{(b) Random select solution} that directly uses samples from different neighbor distributions results in the wrong prediction, because of the heterogeneous distributions; \textbf{(c) Our solution} fits the requests well thanks to our collaborative learning strategy and edge sampling method.}
	\label{intro_pic}
\end{figure*}

Deep Reinforcement Learning (DRL)-based caching policies~\cite{8362276,8374824,8542696} have been verified as possessing the potential to achieve a higher hit rate than the traditional methods. 
However, the DRL agent learns extremely slowly on edge devices operating online due to the sparsity of requests and lack of training time. 
In other words, an edge cache can only collect the local request patterns, insufficiently acquiring the requisite training samples. Furthermore, it is needed to learn a caching policy that is well-adapted to dynamic content popularity (as illustrated in Fig.~\ref{intro_pic}(a)).

The potential for advancing RL caching methods includes using a centralized or multi-agent mechanism, which utilizes request samples collected from all edge caches for faster adaptation and performance improvement~\cite{9155373, 270366}. While such policies are better equipped to adapt to dynamic trends, their usage remains limited due to the heterogeneous distributions on edge devices (illustrated in Fig.~\ref{intro_pic}(b)). Specifically, influenced by such heterogeneity, edge nodes with different distributions are difficult to learn from each other effectively~\cite{8737446, zhou2021caching}.

To address the dynamics issue and the heterogeneous distribution problem, we propose Collaborative Meta Reinforcement Learning Caching with Edge Sampling (CMCES) as illustrated in Fig.~\ref{intro_pic}(c). 
We begin with a meta-reinforcement learning (meta-RL) framework widely used for adapting to the changing distribution of content requests. Our framework creates ``task pairs'' from consecutive time steps and computes a \emph{meta-loss} based on these pairs to reflect the environment's dynamics. 
We reformulate the meta-loss by further considering the cross edges of the ``task pairs''. To increase the sample efficiency, we incorporate a decentralized collaborative mechanism into our framework, so that models learn {\em not only} from their local historical samples, {\em but also} those from their neighboring peers (nodes).

For heterogeneous data distributions, we propose an edge sampling method to make the best use of them. Specifically, we employ an adaptive combination matrix that adjusts the reference weights of other edge nodes, such that the neighbor nodes with similar data distribution as the local node would have maximized weights. This enables a more efficient selection of cooperative nodes and leverages the data across heterogeneous distributions in a sample-efficient way.

Our contribution can be summarized as follows: 

$\rhd$ To address the challenge of dynamic content popularity and improve sample efficiency, we propose a decentralized collaborative edge caching framework augmented with a meta-RL technique.

$\rhd$ To address the issue of heterogeneous distributions, we have developed an edge sampling method for selecting highly valuable cooperative nodes and extracting more beneficial data from those nodes.

$\rhd$ To verify the effectiveness of our solution, we conduct experiments using real-world traces consisting of both long video and short video platform requests. The results show that our design could increase the hit rate by $10.12\%$ (normalized) compared to other baselines.

\section{Related Works}
\noindent\textbf{Deep Reinforcement Learning-Based Caching.}~
Deep reinforcement learning is a powerful framework for designing caching strategies to deal with caching problems without prior assumptions \cite{8362276,8542696}. It has been shown that RL-based methods achieve better hit rates than traditional strategies in practical scenarios \cite{8374824}. Somuyiwa et al.~\cite{8374824} proposed a policy-based method to solve the challenge of unknown data distribution in the content replacement problem. Zhu et al.~\cite{8362276} and Zhong et al.~\cite{8542696} developed actor-critic strategies to leverage better the on-policy way to optimize the agent parameter in limited steps. However, all the above methods cannot quickly adapt to the edge environment because of dynamic request patterns and limited local request history.

\noindent\textbf{Collaborative RL Caching.}~ 
Collaborative caching schemes for edge caches could improve content delivery network performance by frequently exchanging information with neighbors. Wang et al.~\cite{9155373} proposed a multi-agent DRL framework to transmit neighbor nodes’ policy parameters to improve the local model. Bahman et al.~\cite{9596584} designed a key freshness-driven matrix to form the action during the collaborative process. Yan et al.~\cite{270366} developed a knowledge-distilled module to help improve data usage. The limitation of previous studies is that they do not consider the heterogeneous distributions of request patterns among different edge areas, which impairs the effectiveness of collaboration. 

\section{Preliminaries}
In this section, we first present preliminary concepts of reinforcement learning (RL)-based caching, where the cache replacement is modeled as a Markov Decision Process (MDP). Second, we briefly introduce the meta-RL framework for continuous adaptation, a typical paradigm to adapt RL models in dynamic environments.

\subsection{RL-based Caching}
We first model the edge caching problem as an MDP process and introduce the design of the state, action, and reward as follows.

\noindent\textbf{State}: The cache state $s_{n,t}$ of an edge node $n$ includes a list of cache units ($s_{n,t}=\{s_t^1,s_t^2,\cdots,s_t^C\}$), where $s_t^i$ represents the binary state of the $i$-th cache unit ($s_t^i=1$ or $s_t^i=0$). The cache has a finite size  $C$ in an edge server. This is ensured by the condition: $\sum_i s_{i}^{t} \leq C$.

\noindent\textbf{Action}: At each time slot $t$, if the content is requested and found in the local or a neighbor's cache, it is called a hit (denotes as a binary variable $f_{n,t}=1$), otherwise a miss ($f_{n,t}=0$). When ``miss'' happens, the edge cache selects an action $a_t = 1,2,\ldots, C+1$ based on its policy $\pi$; $a_t={C+1}$ will not trigger the content replacement, while $a_t = i$ suggests that caching content in position $i$ will be replaced by the newly requested content item.

\noindent\textbf{Reward}: After a hit request has been found in the local cache or the neighboring cache, the caching policy receives a reward $r_t=\alpha*f_{local,t}+\beta*f_{neighbor,t}$. Our goal is to maximize the hit rate over time. 

In traditional RL, the agent makes  decisions toward optimization goals based on past experience and rewards.
In particular, the agent observes the state $s_{n,t}$ of the task environment $T$ in each time slot $t$ and takes action $a_t$ based on the policy $\pi$ with parameter $\theta$. Then the agent receives the reward $r_t$ and transfers to state $s_{n,t+1}$. We then denote the $H$ episodes' trajectory as a sequence $\tau=(s_{n,t},a_t,r_t,s_{n,t+1})$. The agent then updates a new policy with a cumulative reward loss $L_T(\tau):=-\sum_{t=1}^H r_t$.

At each time slot, the RL model makes decisions in two steps. 1) Interaction step: the agent interacts with the environment and samples trajectories. 2) Training step: the agent computes the loss and updates the policy $\pi$.

The agent with parameter $\theta$ samples $K$ trajectories in the task environment $T$, denoted as $\tau_{T,\theta}^{1:K}$. It is trained by minimizing the objective loss of the RL model:
\begin{equation}\label{final_loss}
\mathbb{L}_T(\tau_\theta^{1:K}):=\frac{1}{K} \sum_{k=1}^K L_T(\tau_\theta^k), \tau_\theta^k \sim P_T(\tau \mid \theta).
\end{equation}

By learning the optimal state-value function, we can reach the optimal policy.

\subsection{Meta-RL for Continuous Adaptation}\label{4.2}
The goal of meta-learning methods is to train a model on various learning tasks to solve a new learning task using only a small number of training samples. Typical meta-learning models~\cite{al2017continuous} learn a meta-initialization $\phi$ by capturing a meta-knowledge from a set of sub-tasks $\{T_{1},T_{2},\ldots,T_{N}\}$ with different data distributions. 

Meta-learning is also a method for continuous adaptation to deal with dynamic environments. A crucial improvement for suiting the continuous adaptation is its focus on capturing the meta-knowledge from the dynamic changes of consequent tasks $\{(T_{1}, T_{2}),(T_{2}, T_{3})\ldots,(T_{N-1}, T_{N})\}$ \cite{al2017continuous}.
Specifically, we regard that these tasks correspond to different dynamics and become sequentially dependent due to environmental changes. Hence, the modified model \cite{al2017continuous} exploits this dependence between consecutive tasks and the meta-learning rule, aiming to adapt to the changing environments. The goal is to minimize the meta-loss for meta-initialization $\phi$, i.e., the average loss over this sequence of task pairs: 
\begin{equation}
    \min_\phi  \sum_{i=1}^{{N}-1} \mathbb{L}_{T_i,T_{i+1}}(\phi).
\end{equation}
where the expression for meta-loss on a pair of consecutive tasks is:
\begin{equation}\label{original loss}
	\mathbb{L}_{T_i,T_{i+1}}(\phi):= \mathbb{E}_{\scaleto{\substack{\tau_{T_{i}, \phi}^{1: K} \sim P_{T_{i}}(\tau \mid \phi)\\ \tau_{T_{i+1}, \theta_i}^{1} \sim P_{T_{i+1}}(\tau \mid \theta_i)}}{22pt}} L_{T_{i+1}}(\tau_{T_{i+1},\theta_i}^1).
\end{equation}
Specifically, the $K$ samples of trajectory $\tau_{T_{i}, \phi}^{1: K}$ come from current task $T_i$, and are used to construct the adapted policy parameter $\theta_i$ that is beneficial for the upcoming task $T_{i+1}$. The meta-loss on a pair of consecutive tasks evaluates the performance of the adapted policy on a trajectory $\tau_{T_{i+1}, \theta_i}^{1}$. Hence, optimizing $\mathbb{L}_{T_i, T_{i+1}}(\phi)$ yields an adaptive update that is optimal concerning the changing rule between tasks $T_i$ and $T_{i+1}$. 

To construct the parameter $\theta_i$ of the adapted policy, we start from the meta-initialization $\phi$ and conduct the gradient decent on trajectories $\tau_{T_i,\phi}^{1:K}$, denoted as: 

\begin{equation}\label{MAML}
\begin{aligned}
        \theta_{i}^{m}=&\left\{\begin{array}{l}
		\scaleto{\phi, m=0,\tau_{T_{i}, \phi}^{1: K} \sim P_{T_{i}}(\tau \mid \phi)}{11pt} \\
		\scaleto{\theta_{i}^{m-1}-\alpha \nabla_{\theta_{i}^{m-1}} \left( \sum\limits^K \gamma^{t} r_{t}\right), \tau_{T_{i}, \theta_{i}^{m-1}}^{1: K}\sim P_{T_{i}}(\tau \mid \theta_{i}^{m-1})}{19pt} \end{array}\right.
    \end{aligned}
\end{equation}
where $M\geq m>0$ and $\theta_i^m$ represent the parameter after $m$-steps gradient decent.

Note that this method only works for the continuous adaptation of single-node, and no previous work combines it with collaborative frameworks.

\section{Collaborative Meta-RL Caching with Edge Sampling}
In this section, we propose our framework of Collaborative Meta-RL Caching with Edge Sampling (CMCES) that consists of two processes. First, we introduce the proposed collaborative meta-RL to capture the meta-knowledge from heterogeneous nodes with temporal dynamic distribution in the meta-pretraining process. Then, we design an edge sampling method to select more valuable collaborative neighbors in the meta-adaptation process.

\subsection{Collaborative Meta-RL}
To solve the problem mentioned in Sec.~\ref{4.2}, the collaborative RL agent continuously adapts the parameter from two sources: 1) The historical data of local requests with temporal dynamic distributions; 2) The historical data of neighboring requests that have heterogeneous distributions.
Hence we consider both the ``local-local task pairs'' and the ``neighbor-local task pairs'' in the meta-learning process.
Notably, the previous meta-RL caching work~\cite{al2017continuous} only makes use of ``local-local task pairs''.

In particular, the loss of ``neighbor-local task pairs'' can be defined by analogizing the loss of ``local-local task pairs'' in Eq.~(\ref{original loss}) as follows:
\begin{align}
	\mathbb{L}_{T_{i,j'},T_{i+1,j}}(\phi)& = \mathbb{E}_{\scaleto{\substack{\tau_{T_{i,j'}, \phi}^{1: K} \sim P_{T_{i}}(\tau \mid \phi)\\ \tau_{T_{i+1,j}, \theta}^{1} \sim P_{T_{i+1,j}}(\tau \mid \theta)}}{19pt}} L_{T_{i+1,j}}(\tau_{T_{i+1,j},\theta}^1). \label{nl_pair}
\end{align}
The indices $i$ represent the index of tasks in the whole sub-task sequence. Indice $j$ and $j'$ represent the index of nodes in which the task occurs. When consecutive tasks occur in the same location, it means that the sequential task pair is a ``local-local task pair'', and when successive tasks occur in different locations, it means that it constitutes a ``neighbor-local task pair''. 

So, it imitates the process of dynamic adaptation from an old neighborhood environment to a new local environment, just like in Fig.~\ref{intro_pic}; the adaptation from neighbor distributions to the local distribution could also help the collaboration. Moreover, this equation measures the ``how much'' model $\theta$ adapted from meta-knowledge $\phi$ using data sampled from neighborhood nodes performed in the new local environment. Minimizing the loss on these ``neighbor-local task pairs'' enhances the transfer ability from the data sampled from neighborhood nodes to a local one.

Due to the above intuition, we reformulate the original meta-loss in Eq.~(\ref{original loss}) by complementing it with the loss of ``neighbor-local task pairs'' in Eq.~(\ref{nl_pair}) as:
\begin{equation}\label{gai total loss}
	\mathbb{L}_{TW}(\phi)= \sum_{i=1}^{N-1} [L_{T_{i,j},T_{i+1,j}}(\phi)+ \lambda   \sum_{j'=1}^E L_{T_{i,j'},T_{i+1,j}}(\phi)].
\end{equation}
where $\lambda\in (0,1)$ is the coefficient of the ``neighbor-local task pairs'', since the data efficiency of ``local-local task pairs'' is more significant than ``neighbor-local task pairs'', resulting in more considerable weight.

We minimize $\mathbb{L}_{TW}(\phi)$ to obtain the global meta-knowledge with a collaborative framework design. This \textbf{Meta Pretraining Process} is summarized in Alg.~\ref{Ag1} as ``Stage 1''.
Specifically, we use a stochastic optimization solver to minimize the meta-loss. First, the local and $E$ neighbor cache nodes sample trajectories $\tau_{T_{i, local},\phi}$ and $\tau_{T_{i, neighbor},\phi}$ using policy parameterized with meta-knowledge $\phi$, respectively. By running multiple gradient descent steps (Eq.~(\ref{MAML})), we get the adapted local policy $\pi_{\theta_i^M}$, and then evaluate the meta-loss of the adapted policy in the upcoming task’s trajectories $\tau_{T_{i+1, local},\theta_i^M}$. With the meta-loss, we adopt one-step gradient descent to update the meta-knowledge. We repeat the whole procedure until the meta-loss is sufficiently small.

\begin{algorithm}[t]
	\caption{Collaborative Meta Edge Sampling Caching}
	\label{Ag1}
\begin{algorithmic}[1] 
		\State {\bfseries Input:} Meta day number $\mathcal{N}$, \  Edge cache Neighbor $E$, \  Task $T_{1,1},\cdots,T_{i,j},\cdots, T_{\mathcal{N},E}$, \ Random matrix $B$.
		\State \textbf{Stage 1: Meta Pretraining Process: }
		\State Randomly initialize $\phi_1$
		\Repeat
		\State Gain all task-pairs ${(T_{i,j},T_{i+1,j})}_{i=1:\mathcal{N}-1,j=1:E}$
		\ForAll{Task-pair data}
		\State Sample traj.$\tau_{T_{i,j}, \phi_i}^{1: K}$ using $\pi_{\phi_i}$
		\State Compute $\theta_{i}^{M}=h(\tau_{T_{i}, \phi_i}^{1: K},\phi_i)$ using Eq.~(\ref{MAML})
		\State Sample traj.$\tau_{T_{i+1,j'}^{1:K},\theta_{i}^{M}}^{1}$ using $\pi_{\theta_{i}^{M}}$
		\EndFor
		\State Update $\phi_{i+1}=\phi_i-\eta \nabla_{\phi} \mathbb{L}_{TW}(\phi)$ using Eq.~(\ref{gai total loss})
		\Until{Convergence to obtain a meta-knowledge $\phi_N^*$}
		\State \textbf{Stage 2: Meta Adaptation Process:}
		\While{New dynamic task $T_{\mathcal{N}+i+1,local}$ coming}
		\State Initialize $\theta_{i}=\phi_{\mathcal{N}}^*$
		\ForAll{Edge cache \  $e_{local}$}
		\State Sample traj.$\tau_{T_{\mathcal{N}+i+1,local}, \Phi_i}^{1: K}$ using $\pi_{\phi_\mathcal{N}}$
		\State Execute \textit{Edge neighbor sampling in Alg.~\ref{Ag2}}
		\EndFor
		\EndWhile
\end{algorithmic}
\end{algorithm}

After this meta pretraining process, we need to perform the \textbf{Meta Adaptation Process}, which is noted as ``Stage 2'' in Alg.~\ref{Ag1}.
A core step in it is an edge sampling strategy, which is introduced next.

\begin{algorithm}[t]
	\caption{Edge Sampling Strategy}
	\label{Ag2}
	\begin{algorithmic}[1] 
	    \State \textbf{Initialize} $0 \leq z \leq \underset{b_{i,j}\neq 0}{min}{\frac{[B^2]_{i,j}}{2b_{i,j}}}$, \ matrix $D=B$
        \State \textbf{Sample Step: Sample neighbor caches for local cache}
        \For{$i=1, \cdots, E-1$}
        \State Sample neighbor cache $i$ with probability $p_{i,e_{local}}$
        \If{node $i$ is not sampled}
        $w_{i,e_{local}} = 0$
        \Else
        \ $d_{i,e_{local}} = \frac{d_{i,e_{local}}}{p_{i,e_{local}}}, w_{i,e_{local}} =d_{i,e_{local}}$
        \EndIf
        \EndFor
        \State \textbf{Combination Step: Update local model with extra info}
        \State Calculate the \textbf{neighbor cache weight} $w_{i,i}$ using Eq.~(\ref{self-weight})
        \State Compute $\theta_{i+1}=h(\tau_{T_{\mathcal{N}+i+1,local}, \Phi_i}^{1: K},\Phi_{i})$
		\State Update \textbf{model of local cache} $e_{local}$ using Eq.~(\ref{combine loss})
	\end{algorithmic}
	\label{Ag}
\end{algorithm}

\subsection{Edge Neighbor Sampling Strategy}
To overcome the heterogeneous problem described in Sec.~\ref{sec:intro}, the sampling strategy is designed to transmit neighbor nodes' samples with reasonable weights.
Our strategy is motivated by the edge sampling technique~\cite{10.5555/3367032.3367126} commonly used in distributed optimization.
Here, we introduce a matrix $W$ for weight calculation across edge nodes.
This weight matrix aims to maximize the weights of the neighbor nodes with similar data distribution as the local node. The matrix $W$ is updated in each iteration round by the feedback, and ensures that the weights of similar data distribution nodes become progressively larger.
This way, one can use the data samples from more relevant nodes to overcome the heterogeneous problem.

Based on the combination matrix $W$, the local cache $i$ samples neighbor cache $j$ to get extra information, and the reference weight is scaled by $1/p_{i,j}$. And for the local cache itself, the reference weight is calculated by:

\begin{equation}\label{self-weight}
	w_{i,i}=1-\sum_{j=1 \backslash i}^E w_{i,j} \ for \ any \  j\in[1,\cdots,i-1,i+1,\cdots,E].
\end{equation}

Since edge caches lack enough appropriate samples, it is difficult to train the local model from random initialization sufficiently. For attaining a comparable performance in the collaboration beginning process when weights are not updated yet, we use matrix $B$ to initialize $D$ to help the local cache ride the dynamic trend~\cite{duchi2011dual}.

We assume $E$ nodes to be affected by a matrix $B$ and its corresponding component $b_{i,j}$. The matrix $B$ satisfies $\sum_{i=1}^n b_{i,j}=1$ and $\sum_{j=1}^n b_{i,j}=1$. Every two nodes $i$ and $j$ are linked by the reference weight $w_{i,j}$ with an appear probability $p_{i,j}=\frac{1}{1+\frac{z}{b_{i,j}}}$. The hyper-parameter $z$ denotes a node-dependent sampling rate, controlled by an upper bound $0 \leq z \leq \underset{b_{i,j}\neq 0}{min}{\frac{[B^2]_{i,j}}{2b_{i,j}}}$ \cite{10.5555/3367032.3367126}.

Our edge neighbor sampling strategy is implemented in Alg.~\ref{Ag2}.
In the \emph{combination step}, each local agent first obtains the samples from the neighbor nodes selected in the sample step and then updates the model with a weighted sum of loss function derivatives:

\begin{equation}\label{combine loss}
	\theta_{local}=\theta_{local}-\eta\sum_{j=1}^E w_{local,j}\cdot\nabla_{\theta_j} \mathbb{L}(\tau_{\theta_j}^{1:K}).
\end{equation}

\section{Evaluation}
\subsection{Dataset}
We use content request records from platforms IQIYI and KuaiShou in 13 days and extract the timestamps and locations of each request. The IQIYI traces contains 53,954,230 requests of 417,077 contents, and the KuaiShou traces contains 528,521,600 requests of 1,746,227 contents. The short video platform KuaiShou has more content requests per unit of time and more videos as possibilities, resulting in more dynamic content popularity.
We randomly select 56 representative edge areas (2.05$km$ $\times$ 2.31$km$ on average). The boundaries of the 56 randomly selected areas are between 116.11802 and 116.45622 for longitude and between 39.987778 and 40.258125 for latitude, which includes the urban and suburban areas of Beijing. In our experiments, we assume that a cache is located at the center of the edge area to serve requests. 

\subsection{Experiment Setup}
The baseline methods include Least Recently Used (LRU), Least Frequently Used (LFU), DRL with manual features (Adaptive-RL) \cite{8362276}, Collaborative Popularity Distillation DRL (LWDRL) \cite{270366} and Multi-agent DRL (MADRL) \cite{9155373}.

The learning rates for meta-learning and meta-adaptation are set as 1e-3 and 2e-4. Besides, we set the default meta-pretraining days as five and did gradient descent three times per adaptation round. The reward $\alpha$, $\beta$, and $\gamma$ are set as 5, 1, and 0.99.

We use the cache hit rate as the evaluation metric. All algorithms execute online. In other words, the model only goes through the dataset once instead of repeatedly learning.

\subsection{Performance Evaluation}
\subsubsection{Local Cache Dynamics Adaptation}

\begin{figure}[!t]
	\centering
\begin{minipage}[t]{0.49\linewidth}
    \centerline{\includegraphics[width=0.99\linewidth]{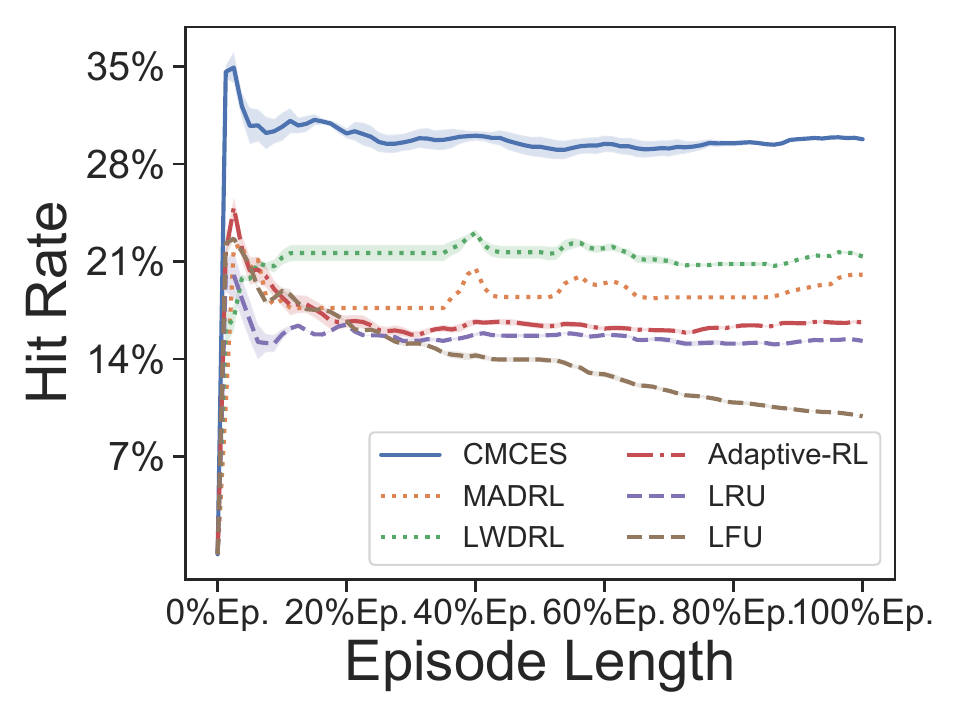}}
\end{minipage}
\medskip
\begin{minipage}[t]{0.49\linewidth}
  \centering
  \centerline{\includegraphics[width=0.99\linewidth]{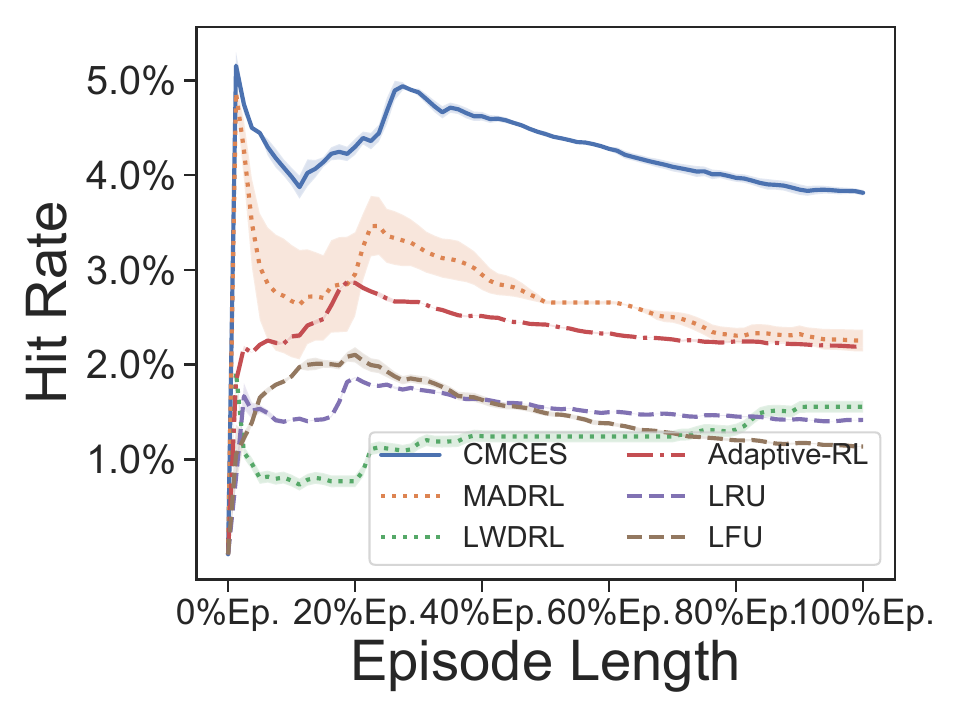}}
\end{minipage}
\vspace{-1.0cm}
	\caption{Average hit rate versus episode time among platform IQIYI (left) and KuaiShou (right).}
	\label{fig1}
\end{figure}

Fig.~\ref{fig1} shows the average hit rate for all selected regions over 13 days with a cache size of 20. Fig.~\ref{fig1}(left) shows the results of the IQIYI dataset, where CMCES works consistently well to ride future dynamic trends. The other DRL-based methods also handle this situation by considering other request samples or features. However, they cannot defeat CMCES for ignoring adaptive collaboration. 

Fig.~\ref{fig1}(right) displays the results for the Kuaishou dataset, where the dynamics of content popularity are more severe and complex. Notably, the gap between the curves of CMCES and other baselines in KuaiShou is more significant than IQIYI in Fig.~\ref{fig1}(left). In this frequently updated environment, CMCES outperforms the other baselines. In contrast, other DRL methods that rely on heterogeneous content distribution samples or outdated distillation knowledge may be inconsistent with current distributions, resulting in worse performance. Our algorithm performs better than the other algorithms, with an average hit rate of 9.3\% (IQIYI) and 1.7\% (KuaiShou).

\begin{figure}[!t]
	\centering
\begin{minipage}[t]{0.47\linewidth}
    \centerline{\includegraphics[width=0.99\linewidth]{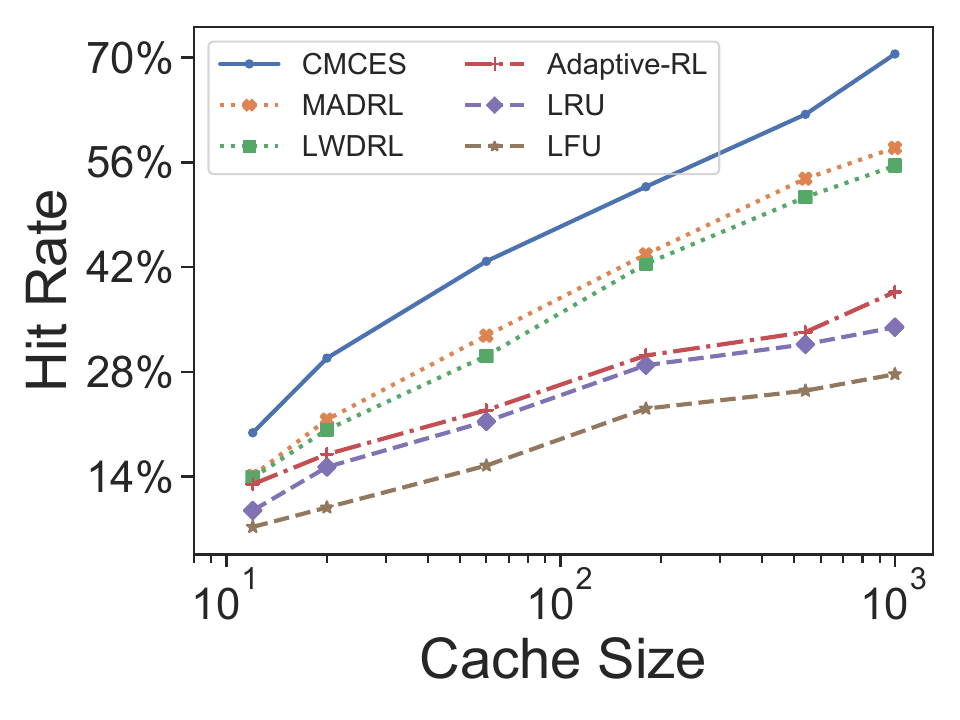}}
\end{minipage}
\medskip
\begin{minipage}[t]{0.47\linewidth}
  \centering
  \centerline{\includegraphics[width=0.99\linewidth]{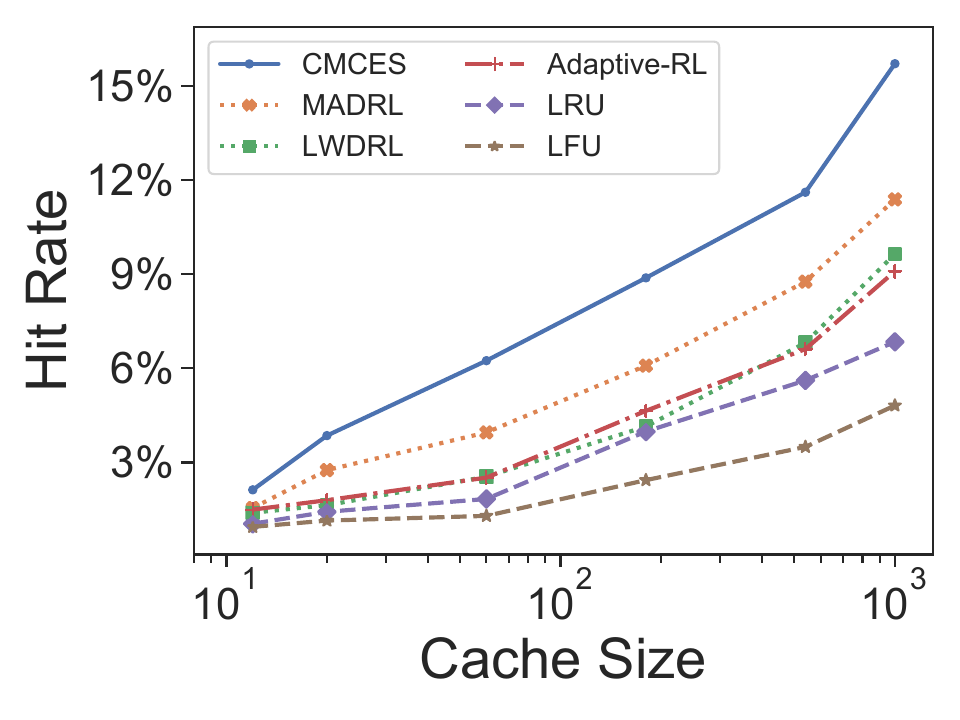}}
\end{minipage}
\vspace{-0.6cm}
	\caption{Average hit rate versus the edge cache size among platform IQIYI (left) and KuaiShou (right).}
	\label{fig2}
\end{figure}
\begin{figure}[!t]
	\centering
\begin{minipage}[b]{0.49\linewidth}
    \centerline{\includegraphics[width=0.99\linewidth]{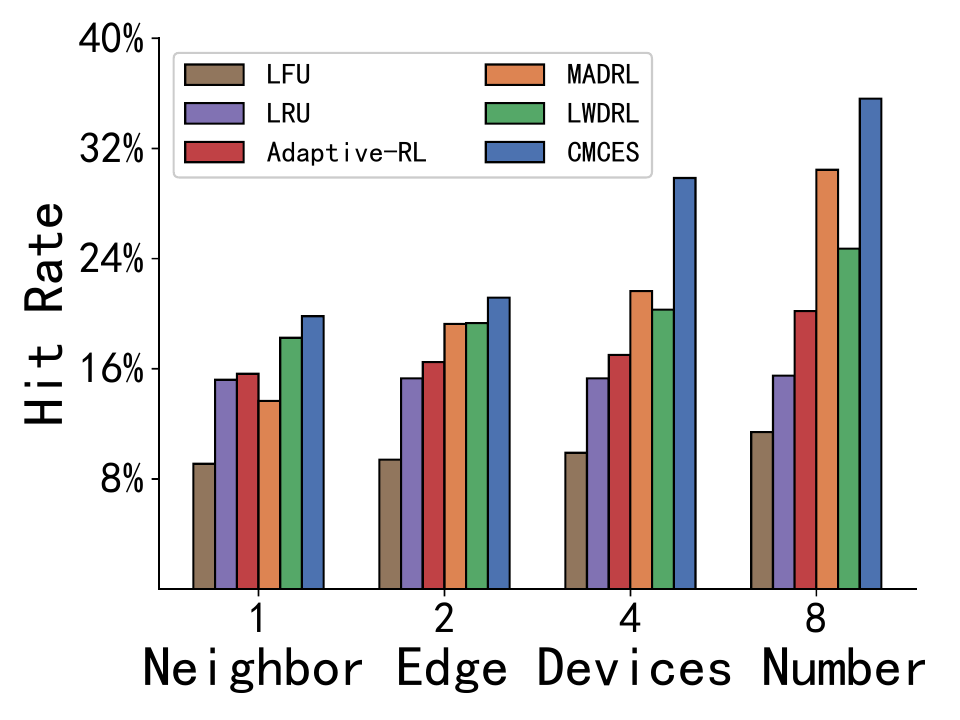}}
\end{minipage}
\medskip
\begin{minipage}[b]{0.49\linewidth}
  \centering
  \centerline{\includegraphics[width=0.99\linewidth]{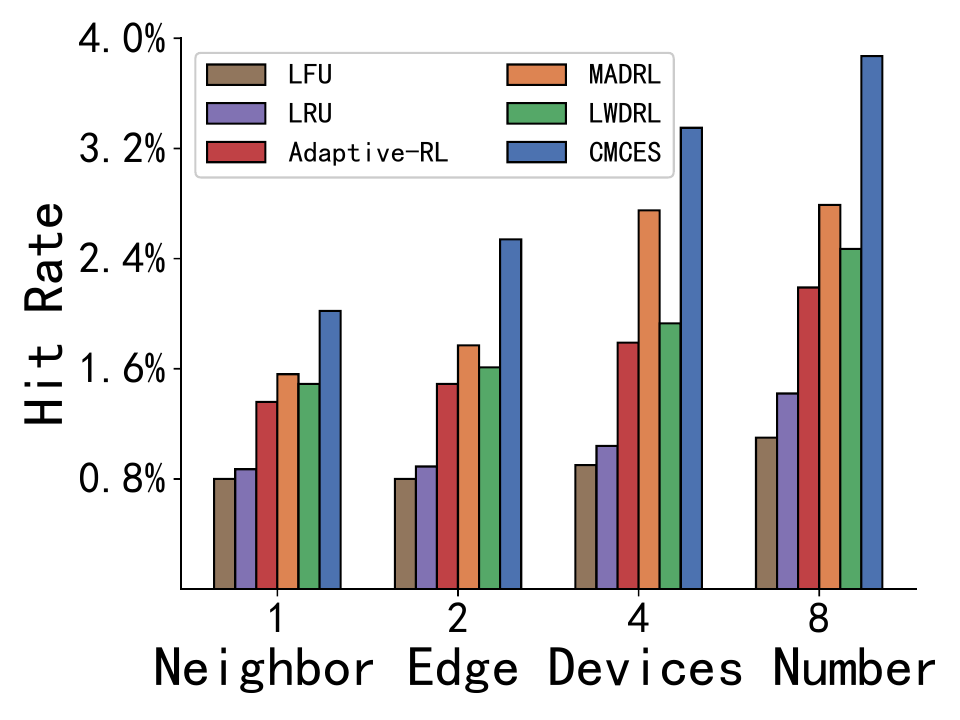}}
\end{minipage}
\vspace{-0.8cm}
	\caption{Average hit rate versus the neighbor cache number among platform IQIYI (left) and KuaiShou (right).}
	\label{fig3}
\end{figure}

\subsubsection{Generality for Different Cache Sizes}
We evaluate the performance of CMCES under the impact of different edge servers’ cache sizes. We experiment on six cache sizes ranging from $10$ to $10^3$. In Fig.~\ref{fig2}, we observe that CMCES outperforms other baselines in all cache sizes. CMCES gains an average improvement of 27.8\% in IQIYI and 42.1\% in Kuaishou than the second rank method. We also observe that the improvement maintains significantly when the cache size increases. The possible reason is that the CMCES learns a better collaborative strategy, when the neighbors have more shared information as the cache size becomes larger.

\subsubsection{Generality of Different Referring Situations}
We consider comparing the performance under different numbers of neighbor edge devices and refer to the 1-neighborhood, 2-neighborhood, 4-neighborhood, and 8-neighborhood situations. Here all cache nodes have 20 cache sizes. As shown in Fig.~\ref{fig3}, CMCES achieves the highest hit rate of all the neighbor numbers for which experiments are conducted, and the hit rate increases as the number of neighbors increases.

\begin{figure}[!t]
	\centering
\begin{minipage}[b]{0.49\linewidth}
    \centerline{\includegraphics[width=0.99\linewidth]{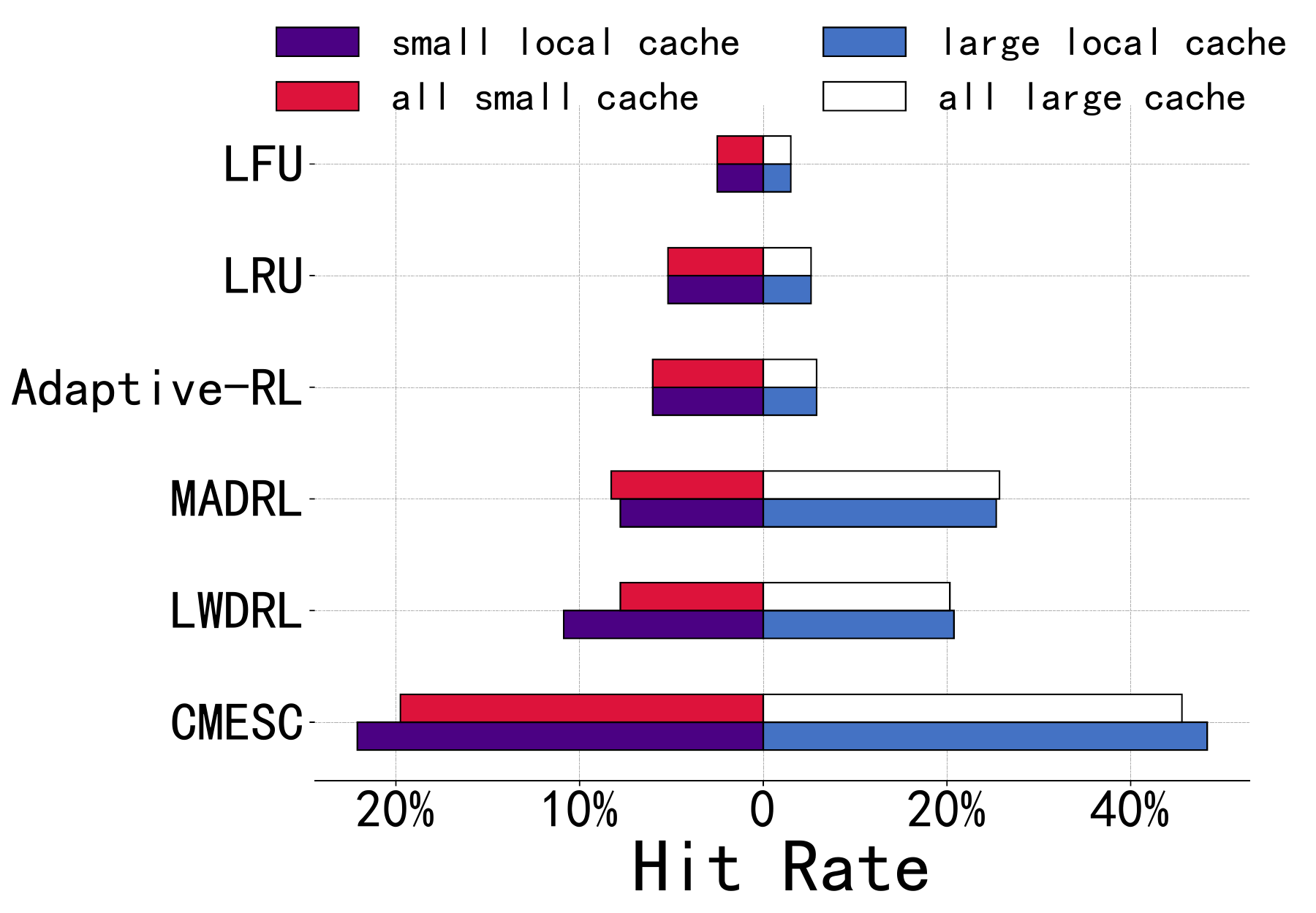}}
\end{minipage}
\medskip
\begin{minipage}[b]{0.49\linewidth}
  \centering
  \centerline{\includegraphics[width=0.99\linewidth]{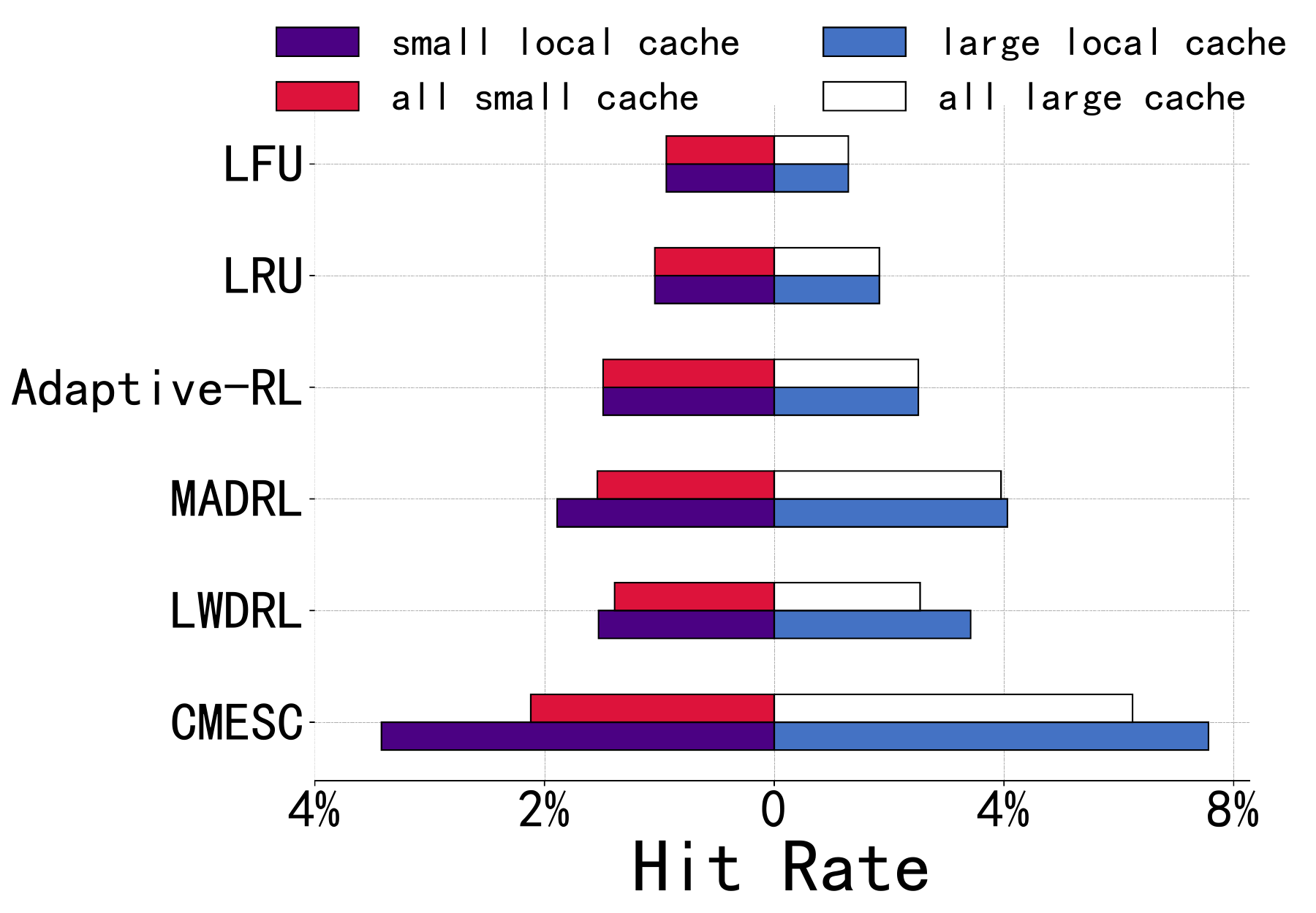}}
\end{minipage}
\vspace{-0.8cm}
	\caption{Average hit rate versus the neighbor cache number among platform IQIYI (left) and KuaiShou (right).}
	\label{fig4}
\end{figure}

Usually, different cache nodes may have different cache sizes in practice. For this case, we measure the performance of small local cache size (100 for local, 600 for neighbor), large local cache size (600 for local, 100 for neighbor), all large cache sizes (600 for all caches), and all small cache size (100 for all caches). Fig.~\ref{fig4} shows that CMCES achieves an average improvement of about 4.51\% and 41.23\% in cache hit rate on long video and short video platforms. Note that when the relative size of the neighbor cache and the local cache is different, the larger neighbor cache provides more improvement to the local cache. The reason is that the larger neighbor cache contains more information on content popularity.

\begin{table}[t]
	\caption{Ablation experiment of different modules}
	\label{tab:Ablation}
        \centering
	\setlength{\tabcolsep}{0.7mm}{
	\begin{tabular}{ccc|cc}
		\hline
		meta-RL& +Collaboration& + Edge Sampling&IQIYI&KuaiShou\\
		\hline
		\hline
		$\bullet$&$\times$&$\times$&18.9\%&2.51\%\\
		$\bullet$&$\bullet$ &$\times$&26.3\%&2.97\%\\
		$\bullet$&$\bullet$&$\bullet$&29.9\%&3.35\%\\
		\hline%
	\end{tabular}}
\end{table}

\subsubsection{Ablation Tests of CMCES}
We ablate the collaborative Meta-RL framework and edge sampling components in turn and look at the changes in hit rate as illustrated in Tab.~\ref{tab:Ablation}.
Compared with the traditional meta-RL, our collaborative meta-RL framework achieves a hit rate increase of 7.4\% (39.1\% improvement) in IQIYI and 0.46\% (18.3\% improvement) in KuaiShou. The reason is that collaboration alleviates the lack of data volume in the local node. Adding the edge sampling module, our CMCES achieves a hit rate increase of 3.6\% (13.7\% improvement) in IQIYI and 0.38\% (12.8\% improvement) in KuaiShou. The results suggest the importance of selecting more valuable collaborative nodes and samples in the case of heterogeneity. The results demonstrate that both collaboration and edge sampling in the proposed method contributes to significant performance improvement.

\subsubsection{Succinctness of Overhead Communication}
We have also measured the communication/transmission cost and overhead package size. Our method reduces the cost by 31\% with a package size of 0.9KB compared to full-volume communication. In comparison, LWDRL~\cite{270366} reduces the cost by 25\% with an overhead package size of 2.0KB in each transmission of popularity distillation from neighbors. MADRL~\cite{9155373} reduces the cost by 26\% in each transmission of the entire model policy parameter from neighbors, with an overhead package size of 58.5KB for $\pi$'s model size. The results demonstrate that the proposed method's transmission cost and overhead package size are outperformance compared to baselines.

\section{Conclusion}
In this paper, we study the problem of collaborative edge caching under continuously changing content popularity. To resolve this, we propose the CMCES framework that enables edge caches to adapt quickly to the local request pattern with neighbors’ exchanging features. In particular, we develop an edge neighbor sampling strategy to mitigate exchanging heterogeneity. The experimental results support our design by the significant outperformance compared to baselines, and demonstrate the efficacy of each component.


\begin{thebibliography}{15}
\bibitem{10.1145/3339825.3391856}
L.~Sun, Y.~Mao, T.~Zong, Y.~Liu, and Y.~Wang, ``Flocking-based live streaming
  of 360-degree video,'' in {\em Proceedings of the 11th ACM Multimedia Systems
  Conference}, p.~26–37, 2020.

\bibitem{7128393}
Y.~Zhou, L.~Chen, C.~Yang, and D.~M. Chiu, ``Video popularity dynamics and its
  implication for replication,'' {\em IEEE Transactions on Multimedia},
  vol.~17, no.~8, pp.~1273--1285, 2015.

\bibitem{10.1145/2517349.2522722}
Q.~Huang, K.~Birman, R.~van Renesse, W.~Lloyd, S.~Kumar, and H.~C. Li, ``An
  analysis of facebook photo caching,'' in {\em Proceedings of the
  Twenty-Fourth ACM Symposium on Operating Systems Principles}, p.~167–181,
  2013.

\bibitem{9148214}
W.~A. Aziz, H.~K. Qureshi, A.~Iqbal, and M.~Lestas, ``Accurate prediction of
  streaming video traffic in tcp/ip networks using dpi and deep learning,'' in
  {\em 2020 International Wireless Communications and Mobile Computing},
  pp.~310--315, 2020.

\bibitem{8374824}
S.~O. Somuyiwa, A.~György, and D.~Gündüz, ``A reinforcement-learning
  approach to proactive caching in wireless networks,'' {\em IEEE Journal on
  Selected Areas in Communications}, vol.~36, no.~6, pp.~1331--1344, 2018.

\bibitem{8542696}
H.~Zhu, Y.~Cao, X.~Wei, W.~Wang, T.~Jiang, and S.~Jin, ``Caching transient data
  for internet of things: A deep reinforcement learning approach,'' {\em IEEE
  Internet of Things Journal}, vol.~6, no.~2, pp.~2074--2083, 2019.

\bibitem{8362276}
C.~Zhong, M.~C. Gursoy, and S.~Velipasalar, ``A deep reinforcement
  learning-based framework for content caching,'' in {\em 2018 Annual
  Conference on Information Sciences and Systems}, pp.~1--6, 2018.

\bibitem{9155373}
F.~Wang, F.~Wang, J.~Liu, R.~Shea, and L.~Sun, ``Intelligent video caching at
  network edge: A multi-agent deep reinforcement learning approach,'' in {\em
  IEEE Conference on Computer Communications}, pp.~2499--2508, 2020.

\bibitem{270366}
H.~Yan, Z.~Chen, Z.~Wang, and W.~Zhu, ``Drl-based collaborative edge content
  replication with popularity distillation,'' in {\em 2021 IEEE International
  Conference on Multimedia and Expo}, pp.~1--6, 2021.

\bibitem{10.5555/3367032.3367126}
C.~Zhang, Q.~Li, and P.~Zhao, ``Decentralized optimization with edge
  sampling,'' in {\em Proceedings of the 28th International Joint Conference on
  Artificial Intelligence}, p.~658–664, 2019.

\bibitem{9596584}
B.~Abolhassani, J.~Tadrous, and A.~Eryilmaz, ``Single vs distributed edge
  caching for dynamic content,'' {\em IEEE/ACM Transactions on Networking},
  vol.~30, no.~2, pp.~669--682, 2022.

\bibitem{al2017continuous}
M.~Al-Shedivat, T.~Bansal, Y.~Burda, I.~Sutskever, I.~Mordatch, and P.~Abbeel,
  ``Continuous adaptation via meta-learning in nonstationary and competitive
  environments,'' {\em International Conference on Learning Representations},
  pp.~1--21, 2018.

\bibitem{8737446}
G.~S. Paschos, A.~Destounis, L.~Vigneri, and G.~Iosifidis, ``Learning to cache
  with no regrets,'' in {\em IEEE INFOCOM 2019 - IEEE Conference on Computer
  Communications}, pp.~235--243, 2019.

\bibitem{zhou2021caching}
S.~Zhou, Z.~Wang, C.~Hu, Y.~Mao, H.~Yan, C.~Wu, S.~Zhang, and W.~Zhu, ``Caching
  in dynamic environments: A near-optimal online learning approach,'' {\em IEEE
  Transactions on Multimedia}, 2021.

\bibitem{duchi2011dual}
J.~C. Duchi, A.~Agarwal, and M.~J. Wainwright, ``Dual averaging for distributed
  optimization: Convergence analysis and network scaling,'' {\em IEEE
  Transactions on Automatic control}, vol.~57, no.~3, pp.~592--606, 2011.
\end{thebibliography}
\end{document}